\documentclass[10pt,twocolumn,twoside]{IEEEtran}
\usepackage{amsmath,array,amssymb,graphicx,color}
\usepackage{cite}
\usepackage{url}
\usepackage{amsthm}
\usepackage {colortbl,array,xcolor}

\newcommand{\mb}[1]{\mathbf{#1}}
\newcommand{\tb}[1]{\textcolor{blue}{#1}}
\newcommand{\tr}[1]{\textcolor{red}{#1}}

\def\thline{\noalign{\hrule height 1pt}}
\def\I{\mathbf{I}}
\def\O{\textbf{O}}
\def\P{\mathbf{P}}
\def\Wu{\boldsymbol{\mathcal{U}}}
\def\Wp{\boldsymbol{\mathcal{P}}}
\def\DWT{\boldsymbol{\mathcal{W}}}
\def\WSST{\boldsymbol{\mathfrak{T}}}
\def\XSTT{\boldsymbol{\mathfrak{S}}}

\begin{document}

\title{
Edge-Aware Extended Star-Tetrix Transforms for\\CFA-Sampled Raw Camera Image Compression
}
\author{
Taizo~Suzuki,~\IEEEmembership{Senior Member,~IEEE}, and Liping~Huang,~\IEEEmembership{Student Member,~IEEE}
\thanks{This work was supported by JSPS Grant-in-Aid for Scientific Research (C) under Grant 22K04084.}
\thanks{T. Suzuki is with the Faculty of Engineering, Information and Systems, University of Tsukuba, Ibaraki 305-8573, Japan (e-mail: taizo@cs.tsukuba.ac.jp).}
\thanks{L. Huang is with Graduate School of Science and Technology, University of Tsukuba, Ibaraki 305-8573, Japan (e-mail: huang@wmp.cs.tsukuba.ac.jp).}
}

\maketitle

\begin{abstract}
Codecs using spectral-spatial transforms efficiently compress raw camera images captured with a color filter array (CFA-sampled raw images) by changing their RGB color space into a decorrelated color space.
This study describes two types of spectral-spatial transform, called extended Star-Tetrix transforms (XSTTs), and their edge-aware versions, called edge-aware XSTTs (EXSTTs), with no extra bits (side information) and little extra complexity.
They are obtained by (i) extending the Star-Tetrix transform (STT), which is one of the latest spectral-spatial transforms, to a new version of our previously proposed wavelet-based spectral-spatial transform and a simpler version, (ii) considering that each 2-D predict step of the wavelet transform is a combination of two 1-D diagonal or horizontal-vertical transforms, and (iii) weighting the transforms along the edge directions in the images.
Compared with XSTTs, the EXSTTs can decorrelate CFA-sampled raw images well: they reduce the difference in energy between the two green components by about $3.38$--$30.08$ \% for high-quality camera images and $8.97$--$14.47$ \% for mobile phone images.
The experiments on JPEG 2000-based lossless and lossy compression of CFA-sampled raw images show better performance than conventional methods.
For high-quality camera images, the XSTTs/EXSTTs produce results equal to or better than the conventional methods: especially for images with many edges, the type-I EXSTT improves them by about $0.03$--$0.19$ bpp in average lossless bitrate and the XSTTs improve them by about $0.16$--$0.96$ dB in average Bj\o ntegaard delta peak signal-to-noise ratio.
For mobile phone images, our previous work perform the best, whereas the XSTTs/EXSTTs show similar trends to the case of high-quality camera images.
\end{abstract}
\section{Introduction}\label{Sec_Intro}
\begin{figure}[t]
		\centering
		\begin{tabular}{c}
			\includegraphics[scale=0.25,keepaspectratio=true]{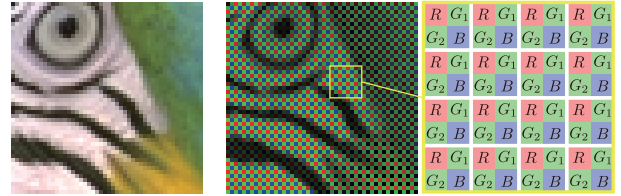}
		\end{tabular}
		\caption{Bayer pattern of a particular area of \textit{Parrot} in the Kodak images dataset~\cite{KODAKTestImages} (each $2\times 2$ pixel square is a macropixel): (left) RGB full-color RGB image and (right) simulated CFA-sampled raw image with corresponding diagram.}
		\label{Fig_CFA}
	\end{figure}

\PARstart{R}{AW} camera images are mainly created by placing a color filter array (CFA) between the light sensors and the camera lens.
To economize on hardware, most cameras capture a color image with a single sensor instead of using three RGB sensors.
In other words, each pixel of such a sensor collects a single color component, either red, green, or blue, not all three, and the obtained raw data is called a CFA-sampled raw image.
The Bayer CFA is the most popular type (Fig.~\ref{Fig_CFA}).
Using a CFA-sampled raw image as is, an image processor performs most of the preprocessing, including black level correction, white balance, demosaicing~\cite{Malvar2004ICASSP,Kiku2016TIP,Tan2018TIP,Wang2020TCSVT}, and gamma correction.
In particular, the image quality largely depends on the performance of the demosaicing process that produces the full-color RGB image we see.
Moreover, performing compression after demosaicing (the demosaicing-first approach) causes redundancy wherein the data volume of a full-color RGB image is three times that of the original CFA-sampled raw image, whereas performing compression before demosaicing (the compression-first approach) can avoid this redundancy.
This compression-first approach also allows us to process images more freely than the demosaicing-first one which almost automatically performs various image processings before compression.
In light of these facts, although many traditional image-compression methods take the demosaicing-first approach, as do standards such as JPEG~\cite{Wallace1992IEEE} and JPEG 2000~\cite{Skodras2001IEEE}, JPEG XS 2nd Edition~\cite{Descampe2021ProcIEEE} takes both approaches.
We believe that the compression-first approach will get more attention in the future from not only professional photographers and designers but also typical consumers.

Spectral-spatial transforms for the compression-first approach have been widely researched~\cite{Zhang2006TIP,Malvar2012DCC,HernandezCabronero2018TMM,Lee2018TIP,Lee2020SPL,Suzuki2020TIP,Richter2021TIP}.
They change a CFA-sampled raw image in RGB color space into data in a decorrelated color space, such as the YDgCbCr or YDgCoCg color space composed of luma, difference green, and two chroma components.
The spectral redundancy between the decorrelated components is very small.
Since the human visual system is not very sensitive to distortion of high-frequency (different green) and chroma components, the strong compression of the decorrelated components will not affect the image quality much.
This study focuses on our previous work on wavelet-based spectral-spatial transforms (WSSTs)~\cite{Suzuki2020TIP} that are represented by cascading (discrete) wavelet transforms, such as Haar, 5/3, and 9/7 wavelet transforms, and cover the other spectral-spatial transforms~\cite{Zhang2006TIP,Malvar2012DCC,HernandezCabronero2018TMM,Lee2018TIP}.
Additionally, there are several new spectral-spatial transforms that are not described in \cite{Suzuki2020TIP}.
One is a newer spectral-spatial transform presented by Lee et al. in \cite{Lee2020SPL} for a CFA-sampled raw image coding framework, called camera-aware multi-resolution analysis (CAMRA)~\cite{Lee2018TIP}.
The YDgCoCg2-WSSTs in \cite{Suzuki2020TIP} generalize the transform in \cite{Lee2020SPL}: the transform in \cite{Lee2020SPL} is obtained by simply applying a 5/3 wavelet transform instead of a Haar transform between the LH and HL subbands to the existing transforms in \cite{Lee2018TIP}.
The other one is the Star-Tetrix transform (STT) of the low-latency low-complexity image coding standard, JPEG XS, developed by Richter et al. \cite{Richter2021TIP}.
It is constructed from three 5/3 wavelet transforms.
The pixels used in the predict and update steps are not limited to be within a macropixel; the surrounding pixels are also used to make the predict results more accurate, as in the case of the existing WSSTs.
On the other hand, for full-color RGB images, (non-redundant) adaptive directional wavelet transforms~\cite{Ding2007TIP,Tanaka2010TIP,Yoshida2011IEICE}, which adapt the filtering directions along the edge information, are efficient and popular methods that take into account image features.
Note that, compared with directional transforms, direct use of wavelet transforms for any of the existing spectral-spatial transforms amounts to nothing but ignoring the image features, especially, edge information.
However, the existing directional transforms do so with extra bits (side information), which adversely affect coding performance, and a significant amount of complexity.

This study develops extended STT (XSTTs) and edge-aware XSTTs (EXSTTs) with no side information and little extra complexity.
They are obtained by (i) extending the STT~\cite{Richter2021TIP} to a new version of the WSSTs~\cite{Suzuki2020TIP} and a simpler version, (ii) considering that each 2-D predict step of the wavelet transforms is a combination of two 1-D diagonal or horizontal-vertical transforms, and (iii) weighting the transforms along the edge directions in the images.
Compared with XSTTs, the EXSTTs can decorrelate CFA-sampled raw images well.
In experiments on JPEG 2000-based lossless and lossy compression of high-quality camera images, our XSTTs/EXSTTs produce results equal to or better than the conventional methods.
Especially for images with many edges, our XSTTs/EXSTTs outperform the conventional methods because of their more efficient decorrelation.
Note that in lossy compression, unlike lossless compression, it is more practical to use XSTTs instead of EXSTTs because the lossy weights are re-calculated and used in the decoder.
In experiments on mobile phone images, which are relatively noisy, our previous work performs the best, whereas the XSTTs/EXSTTs show similar trends to the case of high-quality camera images.

A preliminary version of this study was presented in \cite{Huang2022ICASSP}, where we discussed only the edge-aware weighted 2-D diagonal predict steps for the YDgCoCg-WSSTs in \cite{Suzuki2020TIP}.
In addition to that, this paper extends the STTs in \cite{Richter2021TIP} to XSTTs and further presents the edge-aware weighted 2-D horizontal-vertical predict steps for the XSTTs.

The remainder of the paper is organized as follows.
Section~\ref{Sec_Review} reviews the conventional methods: WSSTs and STT.
Section~\ref{Sec_Prop} extends the STT to two new versions of the WSSTs, i.e., the XSTTs, and introduces two types of edge-aware weighted 2-D predict steps to be applied to the XSTTs.
Section~\ref{Sec_Result} shows the effect of the edge-aware weighted 2-D predict steps and compares the resulting XSTTs/EXSTTs with the conventional methods in JPEG 2000 for CFA-sampled raw image compression.
Section~\ref{Sec_Concl} concludes the paper.

\textit{Notation}: Boldface letters represent vectors and matrices.
$\I$ and $\O$ denote a $2\times 2$ identity matrix and zero matrix, respectively.
Moreover, $\cdot^\top$, $\lfloor\cdot\rfloor$, and $|\cdot|$ denote the transpose, floor function (rounding operation), and absolute value, respectively.
Let $z_i$ be a horizontal ($i=1$) or vertical ($i=2$) delay element and $\overline{z}_i=z_i^{-1}$.
In addition, the size and dynamic range of the images in the figures have been adjusted for display.

\section{Review and Definitions}\label{Sec_Review}
\subsection{Wavelet-Based Spectral-Spatial Transforms}
	\begin{table}[t]
		\centering
		\caption{Coefficients of 5/3 and 9/7 wavelet transforms.}
		\begin{tabular}{c|cc}
		\thline
		& 5/3 & 9/7 \\
		\hline
		$p_0$ & $-1/2$ & $-1.58613434205992$ \\
		$u_0$ & $1/4$  & $-0.05298011857295$ \\
		$p_1$ & $0$    & $0.882911075530940$ \\
		$u_1$ & $0$    & $0.443506852043967$ \\
		\thline
		\end{tabular}
		\label{Tab_Coef}
	\end{table}
	\begin{figure*}[t]
		\centering
		\includegraphics[scale=0.3,keepaspectratio=true]{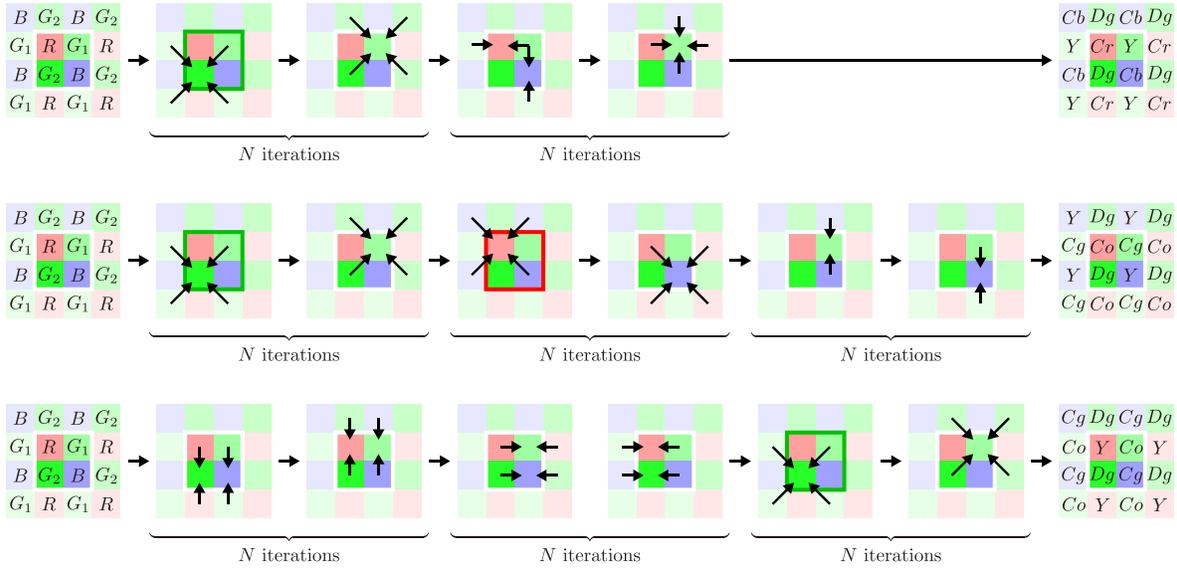}
		\caption{Implementations of WSSTs (arrows, green-framed macropixels, and red-framed macropixels mean lifting step, $G1$-to-$G2$ prediction, and $B$-to-$R$ prediction, respectively): (top-to-bottom) YDgCbCr-, YDgCoCg-, and YDgCoCg2-WSSTs.}
		\label{Fig_Implements_WSST}
	\end{figure*}
Our previous work~\cite{Suzuki2020TIP} presented WSSTs that cover many spectral-spatial transforms for CFA-sampled raw image compression.
The WSST $\WSST$ is represented as
	\begin{align}
		\begin{bmatrix}
			Y, D_\text{g}, C_1, C_2
		\end{bmatrix}^\top
		=
		\WSST
		\begin{bmatrix}
			G_1, G_2, B, R
		\end{bmatrix}^\top
		,
	\end{align}
where $R$, $G_1$, $G_2$, $B$, $Y$, $D_\text{g}$, $C_1$, and $C_2$ mean red, green, other green, blue, luma, difference green, chroma, and other chroma components, respectively.
The $\WSST$s are classified into three types: YDgCbCr-WSSTs $\WSST_\text{br}$, YDgCoCg-WSSTs $\WSST_\text{og}$, and YDgCoCg2-WSSTs $\WSST_\text{og2}$, as follows:
	\begin{align}
		\WSST_\text{br} 
		=&~
		\P_0
		\begin{bmatrix}
			1  & \O \\
			\O & \DWT_3(\overline{z}_1,z_2)
		\end{bmatrix}
		\P_0
		\begin{bmatrix}
			\DWT_2(\overline{z}_1,z_2) & \O \\
			\O                 & \I
		\end{bmatrix}
		,
		\\
		\WSST_\text{og} 
		=&~
		\P_2
		\begin{bmatrix}
			\DWT_2(\overline{z}_2) & \O \\
			\O             & \I
		\end{bmatrix}
		\P_1
		\begin{bmatrix}
			\DWT_2(\overline{z}_1,z_2) & \O \\
			\O                 & \DWT_2(\overline{z}_1,\overline{z}_2)
		\end{bmatrix}
		,
		\\
		\WSST_\text{og2} 
		=&~
		\P_5
		\begin{bmatrix}
			\DWT_2(\overline{z}_1,z_2) & \O \\
			\O                         & \I
		\end{bmatrix}
		\P_4
		\begin{bmatrix}
			\DWT_2(z_1) & \O \\
			\O          & \DWT_2(z_1)
		\end{bmatrix}
		\nonumber
		\\
		&~
		\cdot
		\P_3
		\begin{bmatrix}
			\DWT_2(z_2) & \O \\
			\O          & \DWT_2(z_2)
		\end{bmatrix}
		\P_2
		,
	\end{align}
where $\DWT_2(z_i)$, $\DWT_2(z_1,z_2)$, and $\DWT_3(z_1,z_2)$ are wavelet transforms:
	\begin{align}
		\DWT_2(z_i)
		=&~
		\prod_{k=N-1}^{0}
		\underbrace{
		\begin{bmatrix}
			1 & U_k(z_i) \\
			0 & 1
		\end{bmatrix}
		}_{\text{update step}}
		\underbrace{
		\begin{bmatrix}
			1        & 0 \\
			P_k(z_i) & 1
		\end{bmatrix}
		}_{\text{predict step}}
		,
		\label{Eqs_DWT1}
		\\
		\DWT_2(z_1,z_2)
		=&~
		\prod_{k=N-1}^{0}
		\underbrace{
		\begin{bmatrix}
			1 & U_k(z_1,z_2) \\
			0 & 1
		\end{bmatrix}
		}_{\text{update step}}
		\underbrace{
		\begin{bmatrix}
			1            & 0 \\
			P_k(z_1,z_2) & 1
		\end{bmatrix}
		}_{\text{predict step}}
		,
		\label{Eqs_DWT2}
		\\
		\DWT_3(z_1,z_2)
		=&~
		\prod_{k=N-1}^{0}
		\underbrace{
		\begin{bmatrix}
			1        & \O \\
			\frac{1}{2}
			\begin{bmatrix}
				U_k(z_1) \\
				U_k(z_2)
			\end{bmatrix}
			& \I
		\end{bmatrix}
		^\top
		}_{\text{update step}}
		\underbrace{
		\begin{bmatrix}
			1        & \O \\
			\begin{bmatrix}
				P_k(z_1) \\
				P_k(z_2)
			\end{bmatrix}
			& \I
		\end{bmatrix}
		}_{\text{predict step}}
		.
		\label{Eqs_DWT3}
	\end{align}
$\P_j$ ($j\in\mathbb{N}$) is a $4\times 4$ permutation matrix,
	\begin{align}
		\P_0
		&
		=
		\begin{bmatrix}
			0 & 1 & 0 & 0 \\
			1 & 0 & 0 & 0 \\
			0 & 0 & 1 & 0 \\
			0 & 0 & 0 & 1 \\
		\end{bmatrix}
		,\ 
		\P_1
		=
		\begin{bmatrix}
			0 & 0 & 1 & 0 \\
			1 & 0 & 0 & 0 \\
			0 & 1 & 0 & 0 \\
			0 & 0 & 0 & 1 \\
		\end{bmatrix}
		,
		\nonumber
		\\
		\P_2
		&
		=
		\begin{bmatrix}
			1 & 0 & 0 & 0 \\
			0 & 0 & 1 & 0 \\
			0 & 0 & 0 & 1 \\
			0 & 1 & 0 & 0 \\
		\end{bmatrix}
		,\ 
		\P_3
		=
		\begin{bmatrix}
			0 & 0 & 1 & 0 \\
			1 & 0 & 0 & 0 \\
			0 & 0 & 0 & 1 \\
			0 & 1 & 0 & 0 \\
		\end{bmatrix}
		,
		\nonumber
		\\
		\P_4
		&
		=
		\begin{bmatrix}
			0 & 1 & 0 & 0 \\
			0 & 0 & 1 & 0 \\
			1 & 0 & 0 & 0 \\
			0 & 0 & 0 & 1 \\
		\end{bmatrix}
		,\ 
		\P_5
		=
		\begin{bmatrix}
			0 & 0 & 1 & 0 \\
			0 & 1 & 0 & 0 \\
			1 & 0 & 0 & 0 \\
			0 & 0 & 0 & 1 \\
		\end{bmatrix}
		,
	\end{align}
and $P_k(z_i)$, $U_k(z_i)$, $P_k(z_1,z_2)$, and $U_k(z_1,z_2)$ are polynomials with coefficients $p_k$ and $u_k$:
	\begin{align}
		P_k(z_i)     &= (1+\overline{z}_i)p_k,
		\label{Eqs_1DP}
		\\
		U_k(z_i)     &= (1+z_i)u_k,
		\\
		P_k(z_1,z_2) &= \frac{1}{2}(1+\overline{z}_1+\overline{z}_2+\overline{z}_1\overline{z}_2)p_k,
		\label{Eqs_2DP}
		\\
		U_k(z_1,z_2) &= \frac{1}{2}(1+z_1+z_2+z_1z_2)u_k
		.
	\end{align}
Table~\ref{Tab_Coef} shows the coefficients $p_k$ and $u_k$ in the 5/3 wavelet transforms ($N=1$) and 9/7 wavelet transforms ($N=2$), and Fig.~\ref{Fig_Implements_WSST} shows implementations of the existing WSSTs.
The pixels used in the predict and update steps are not limited to be within a macropixel; the surrounding pixels are also used to make the predict results more accurate.
However, because they do not consider image features, their predictions may not be so accurate in some cases.

\subsection{Star-Tetrix Transform}
The STT, presented by Richter et al.~\cite{Richter2021TIP}, is a spectral-spatial transform for CFA-sampled raw image compression with JPEG XS.
It consists of four steps based on 5/3 wavelet transforms.
The first step generates chroma components $C_\text{b}$ and $C_\text{r}$ by predicting $R$ and $B$ from the four surrounding green components $G^x$ ($x=\{l,r,t,b\}$), where $l$, $r$, $t$, and $b$ respectively indicate the sample position to the left, right, top, and bottom of the current pixel, as
	\begin{align}
		C_\text{b}
		&=
		B-\left\lfloor\frac{G^l+G^r+G^t+G^b}{4}\right\rfloor
		,
		\label{Eqs_STT_Cb}
		\\
		C_\text{r}
		&=
		R-\left\lfloor\frac{G^l+G^r+G^t+G^b}{4}\right\rfloor
		.
		\label{Eqs_STT_Cr}
	\end{align}
The second step generates the luma components $Y_1$ and $Y_2$ by updating $G$ from the four surrounding $C_\text{b}^x$s and $C_\text{r}^x$s as
	\begin{align}
		Y_1
		&=
		G+\left\lfloor\frac{C_\text{r}^l+C_\text{r}^r+C_\text{b}^t+C_\text{b}^b}{8}\right\rfloor
		,
		\label{Eqs_STT_Y1}
		\\
		Y_2
		&=
		G+\left\lfloor\frac{C_\text{r}^t+C_\text{r}^b+C_\text{b}^l+C_\text{b}^r}{8}\right\rfloor
		.
		\label{Eqs_STT_Y2}
	\end{align}
For simplicity, the white-balancing constants defined in \cite{Richter2021TIP} will not be considered in this study.
The third step generates the luma difference component $\Delta$ by predicting $Y_1$ from the four surrounding $Y_2$s, as
	\begin{align}
		\Delta
		&=
		Y_1-\left\lfloor\frac{Y_2^{l,t}+Y_2^{r,t}+Y_2^{l,b}+Y_2^{r,b}}{4}\right\rfloor
		.
		\label{Eqs_STT_Delta}
	\end{align}
The last step generates the final luma component $Y$ by updating $Y_2$ from the four surrounding $\Delta$s, as
	\begin{align}
		Y
		&=
		Y_2+\left\lfloor\frac{\Delta^{l,t}+\Delta^{r,t}+\Delta^{l,b}+\Delta^{r,b}}{8}\right\rfloor
		.
		\label{Eqs_STT_Y}
	\end{align}
Although the STT here looks different from the WSSTs, we can extend (generalize) the STT to the WSSTs as described in the next section.

\section{Extended Star-Tetrix Transforms and Edge-Aware Weighted 2-D Predict Steps}\label{Sec_Prop}
\subsection{Type-I Extended Star-Tetrix Transforms: A New Wavelet-based Spectral-Spatial Transform}
	\begin{figure*}[t]
		\centering
		\includegraphics[scale=0.3,keepaspectratio=true]{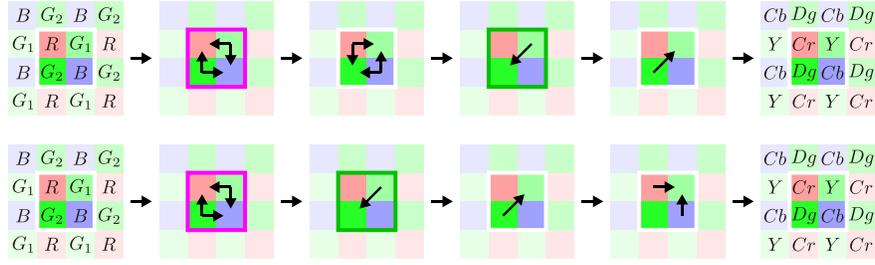}
		\caption{Implementations of XSTTs that use Haar transforms (arrows, green-framed macropixels, and magenta-framed macropixels mean lifting step, $G1$-to-$G2$ prediction, and $(G_1,G_2)$-to-$R$ and $(G_1,G_2)$-to-$B$ predictions, respectively): (top) XSTT-Is and (bottom) XSTT-IIs.}
		\label{Fig_Implement_XSTT_Haar}
	\end{figure*}
\begin{figure*}[t]
		\centering
		\includegraphics[scale=0.3,keepaspectratio=true]{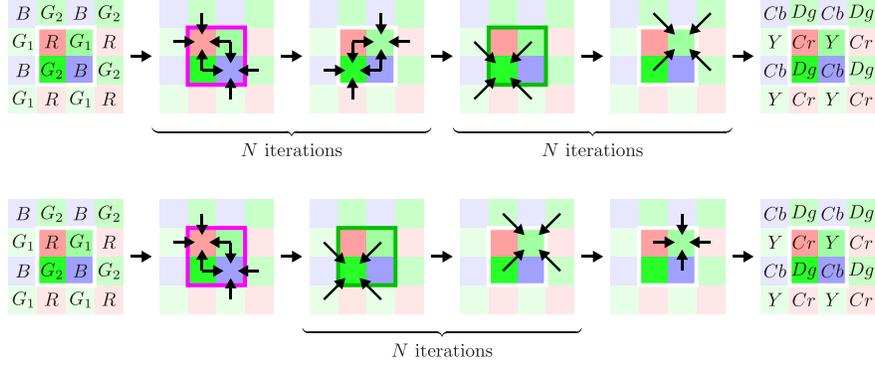}
		\caption{Implementations of XSTTs (arrows, green-framed macropixels, and magenta-framed macropixels mean lifting step, $G1$-to-$G2$ prediction, and $(G_1,G_2)$-to-$R$ and $(G_1,G_2)$-to-$B$ predictions, respectively): (top) XSTT-Is and (bottom) XSTT-IIs.}
		\label{Fig_Implement_XSTT}
	\end{figure*}

\textit{Definition:}
Although the original STT~\cite{Richter2021TIP} consists of three 5/3 wavelet transforms as described above, we can extend it to a fourth version of the WSSTs (second version of the YDgCbCr-WSSTs).
Since another type of extended STT (XSTT) will be introduced in the next subsection, we will refer to the XSTTs in this subsection as ``type-I XSTTs'' or ``XSTT-Is'' to distinguish between them.
The XSTT-I $\XSTT_\text{I}$ is represented as follows:
	\begin{align}
		\XSTT_\text{I}
		=
		\begin{bmatrix}
			\DWT_2(\overline{z}_1,z_2) & \O \\
			\O                         & \I
		\end{bmatrix}
		\prod_{k=N-1}^{0}
		\underbrace{
		\begin{bmatrix}
			\I & \Wu_k \\
			\O & \I    \\
		\end{bmatrix}
		}_{\text{update step}}
		\underbrace{
		\begin{bmatrix}
			\I    & \O \\
			\Wp_k & \I
		\end{bmatrix}
		}_{\text{predict step}}
		\label{Eqs_XSTT1}
		,
	\end{align}
where
	\begin{align}
		\Wp_k
		=
		\frac{1}{2}
		\begin{bmatrix}
			P_k(z_2)            & P_k(z_1) \\
			P_k(\overline{z}_1) & P_k(\overline{z}_2)
		\end{bmatrix}
		,~
		\Wu_k
		=
		\frac{1}{2}
		\begin{bmatrix}
			U_k(z_2) & U_k(\overline{z}_1) \\
			U_k(z_1) & U_k(\overline{z}_2)
		\end{bmatrix}
		\label{Eqs_XSTT1_PU}
		.
	\end{align}
Although we have omitted the white-balancing constants defined in \cite{Richter2021TIP}, the parameters can be simply applied to $\DWT_2(\overline{z}_1,z_2)$ in \eqref{Eqs_XSTT1} if we desire it.
The top of Fig.~\ref{Fig_Implement_XSTT} shows an implementation of the XSTT-Is.
The XSTT-Is will be further extended to edge-aware transforms later.

\textit{Remark-1:}
In the case of the 5/3 wavelet transforms, the first, second, and third steps in \eqref{Eqs_XSTT1} represent the same process as \eqref{Eqs_STT_Cb}-\eqref{Eqs_STT_Cr}, \eqref{Eqs_STT_Y1}-\eqref{Eqs_STT_Y2}, and \eqref{Eqs_STT_Delta}-\eqref{Eqs_STT_Y}, respectively; i.e., the XSTT-I that uses 5/3 wavelet transforms is identical to the original STT.
In addition, $D_\text{g}$ in the XSTT-I is denoted as $\Delta$ in the original STT.

\textit{Remark-2:}
If there are no rounding operations in the lifting steps, we find that the XSTT-I that uses Haar transforms is identical to the YDgCbCr-WSST that uses Haar transforms:
	\begin{align}
		\XSTT_\text{I}
		=&
		\begin{bmatrix}
			1/4  & 1/4  & 1/4 & 1/4 \\
			-1   & 1    & 0   & 0   \\
			-1/2 & -1/2 & 1   & 0   \\
			-1/2 & -1/2 & 0   & 1   \\
		\end{bmatrix}
		\nonumber
		\\
		=&
		\begin{bmatrix}
			1 & 1/2 & 0 & 0 \\
			0 & 1   & 0 & 0 \\
			0 & 0   & 1 & 0 \\
			0 & 0   & 0 & 1 \\
		\end{bmatrix}
		\begin{bmatrix}
			1  & 0 & 0 & 0 \\
			-1 & 1 & 0 & 0 \\
			0  & 0 & 1 & 0 \\
			0  & 0 & 0 & 1 \\
		\end{bmatrix}
		\nonumber
		\\
		&\cdot
		\begin{bmatrix}
			1 & 0 & 1/4 & 1/4 \\
			0 & 1 & 1/4 & 1/4 \\
			0 & 0 & 1   & 0   \\
			0 & 0 & 0   & 1   \\
		\end{bmatrix}
		\begin{bmatrix}
			1    & 0    & 0 & 0 \\
			0    & 1    & 0 & 0 \\
			-1/2 & -1/2 & 1 & 0 \\
			-1/2 & -1/2 & 0 & 1 \\
		\end{bmatrix}
		.
		\label{Eqs_XSTT1_Haar}
	\end{align}
The implementation is shown at the top of Fig.~\ref{Fig_Implement_XSTT_Haar}.
Consequently, the XSTT-Is can be considered to be a second version of the YDgCbCr-WSSTs.
Note that the XSTT-I that uses Haar transforms outputs slightly different results from the YDgCbCr-WSST that uses Haar transforms due to the rounding operations on the lifting steps in the actual implementation.

\subsection{Type-II Extended Star-Tetrix Transforms: A Simpler Version than Type-I Extended Star-Tetrix Transforms}
\textit{Definition:}
Here, we describe another type of XSTT; called the ``type-II XSTTs'' or ``XSTT-IIs.''
These are simpler than the XSTT-Is and a fifth version of the WSSTs (third version of the YDgCbCr-WSSTs).
The XSTT-II $\XSTT_\text{I\hspace{-.1em}I}$ is represented as follows:

	\begin{align}
		\XSTT_\text{I\hspace{-.1em}I}
		=&
		\underbrace{
		\begin{bmatrix}
			\I & \widetilde{\Wu}_0 \\
			\O & \I
		\end{bmatrix}
		}_{\text{update step}}
		\begin{bmatrix}
			\DWT_2(\overline{z}_1,z_2) & \O \\
			\O                         & \I
		\end{bmatrix}
		\underbrace{
		\begin{bmatrix}
			\I    & \O \\
			\Wp_0 & \I
		\end{bmatrix}
		}_{\text{predict step}}
		,
		\label{Eqs_XSTT2}
	\end{align}
where
	\begin{align}
		\widetilde{\Wu}_0
		=
		\frac{1}{2}
		\begin{bmatrix}
			U_0(z_2) & U_0(\overline{z}_1) \\
			0        & 0
		\end{bmatrix}
		\label{Eqs_XSTT2_PU}
		.
	\end{align}
The XSTT-IIs are considered to be special cases of WSSTs because wavelet transforms with $N\geq 2$, such as 9/7 wavelet transforms, can only be applied to the second step in \eqref{Eqs_XSTT2}.
The bottom of Fig.~\ref{Fig_Implement_XSTT} shows an implementation of the XSTT-IIs.
As with the XSTT-Is, they will be further extended to edge-aware transforms later.

\textit{Derivation:}
The XSTT-II in \eqref{Eqs_XSTT2} is obtained with the following simple procedure.
First, the XSTT-I that uses Haar transforms in \eqref{Eqs_XSTT1_Haar} is rewritten as
	\begin{align}
		\XSTT_\text{I\hspace{-.1em}I}
		=&
		\begin{bmatrix}
			1/4  & 1/4  & 1/4 & 1/4 \\
			-1   & 1    & 0   & 0   \\
			-1/2 & -1/2 & 1   & 0   \\
			-1/2 & -1/2 & 0   & 1   \\
		\end{bmatrix}
		\nonumber
		\\
		=&
		\begin{bmatrix}
			1 & 0 & 1/4 & 1/4 \\
			0 & 1 & 0   & 0   \\
			0 & 0 & 1   & 0   \\
			0 & 0 & 0   & 1   \\
		\end{bmatrix}
		\begin{bmatrix}
			1 & 1/2 & 0 & 0 \\
			0 & 1   & 0 & 0 \\
			0 & 0   & 1 & 0 \\
			0 & 0   & 0 & 1 \\
		\end{bmatrix}
		\nonumber
		\\
		&\cdot
		\begin{bmatrix}
			1  & 0 & 0 & 0 \\
			-1 & 1 & 0 & 0 \\
			0  & 0 & 1 & 0 \\
			0  & 0 & 0 & 1 \\
		\end{bmatrix}
		\begin{bmatrix}
			1    & 0    & 0 & 0 \\
			0    & 1    & 0 & 0 \\
			-1/2 & -1/2 & 1 & 0 \\
			-1/2 & -1/2 & 0 & 1 \\
		\end{bmatrix}
		.
		\label{Eqs_XSTT2_Haar}
	\end{align}
Let $\XSTT_\text{I\hspace{-.1em}I}$ be the XSTT-II that uses Haar transforms.
Next, in accordance with the extension method described in \cite{Suzuki2020TIP}, the XSTT-II that uses Haar transforms in \eqref{Eqs_XSTT2_Haar} can easily be extended to WSSTs in the form of \eqref{Eqs_XSTT2}.

\textit{Remark-1:}
The XSTT-IIs are simpler than the XSTT-Is because $\widetilde{\Wu}_0$ in \eqref{Eqs_XSTT2_PU} does not require $U_0(z_1)$ or $U_0(\overline{z}_2)$ belonging to the bottom row of $\Wu_0$ in \eqref{Eqs_XSTT1_PU}, i.e., $U_0(z_1)=U_0(\overline{z}_2)=0$.

\textit{Remark-2:}
Like the XSTT-I, if there are no rounding operations in the lifting steps, the XSTT-II that uses Haar transforms is identical to the YDgCbCr-WSST that uses Haar transforms, as shown in \eqref{Eqs_XSTT2_Haar}.
The implementation is shown at the bottom of Fig.~\ref{Fig_Implement_XSTT_Haar}.
Consequently, the XSTT-IIs can be considered to be a third version of the YDgCbCr-WSSTs.
Note that the XSTT-II that uses Haar transforms outputs slightly different results from the XSTT-I and YDgCbCr-WSST that use Haar transforms due to the rounding operations.

\subsection{Edge-Aware Weighted 2-D Diagonal Predict Steps}
	\begin{figure}[t]
		\centering
		\includegraphics[scale=0.3,keepaspectratio=true]{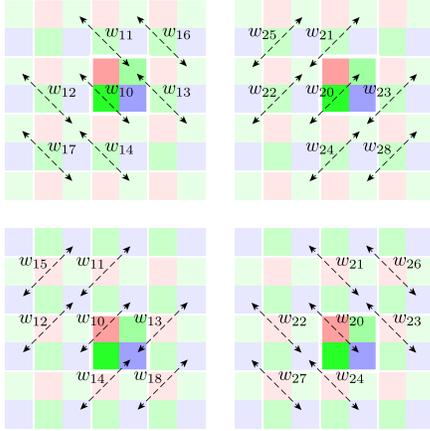}
		\caption{Weights for the 2-D diagonal predict step (dashed arrows mean to calculate the difference between two pixels):
		(top) for $G_1$-to-$G_2$ prediction
		and
		(bottom) $B$-to-$R$ prediction.}
		\label{Fig_WeightsDiag}
	\end{figure}
	\begin{figure}[t]
		\centering
		\includegraphics[scale=0.3,keepaspectratio=true]{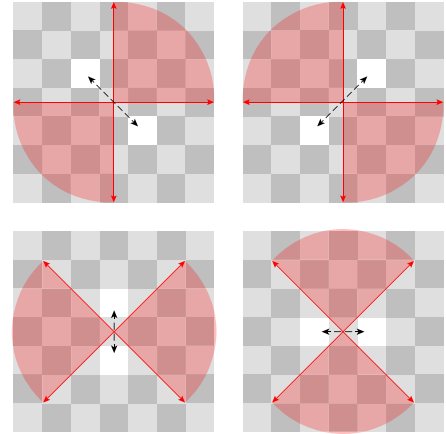}
		\caption{Relation between weight calculation and edge direction detection (dashed arrows mean the difference between two pixels and the red areas are the detectable edge direction ranges):
		(top-left) left diagonal weight,
		(top-right) right diagonal weight,
		(bottom-left) vertical weight,
		and
		(bottom-right) horizontal weight.}
		\label{Fig_Weights2Direction}
	\end{figure}
In this subsection, we extend the polynomial $P_k(z_1,z_2)$ in \eqref{Eqs_2DP}, which is in the 2-D \textit{diagonal} predict step of the wavelet transform $\DWT_2(z_1,z_2)$ in \eqref{Eqs_DWT2}, to
	\begin{align}
		\widetilde{P}_k(z_1,z_2)
		=
		\frac{W_1(1+\overline{z}_1\overline{z}_2)+W_2(\overline{z}_1+\overline{z}_2)}{W_1+W_2}
		p_k
		,
		\label{Eqs_Pk_Diag}
	\end{align}
where $W_m$ ($m=1,2$) is a weight,
	\begin{align}
		W_m
		=
		\sum_{n}w_{mn}+\varepsilon
		\label{Eqs_weight},
	\end{align}
$w_{mn}$ is automatically determined in accordance with the image features as described later, and an extremely small value $\varepsilon$ is added to \eqref{Eqs_weight} to avoid division by zero.
Hence, the original 2-D diagonal predict step is divided into two 1-D diagonal (left- and right-diagonal) predict steps and they are weighted to consider the image features.
When $W_1=W_2=1$, it is clear that $\widetilde{P}_k(z_1,z_2)=P_k(z_1,z_2)$.
Consequently, for the cases of $G_1$-to-$G_2$ and $B$-to-$R$ predictions, we can rewrite $\widetilde{P}_k(z_1,z_2)$ in \eqref{Eqs_Pk_Diag}, as
	\begin{align}
		\widetilde{P}_k(\overline{z}_1,z_2)
		&=
		\frac{W_1(1+z_1\overline{z}_2)+W_2(z_1+\overline{z}_2)}{W_1+W_2}
		p_k
		\nonumber
		\\
		&\qquad\qquad\qquad\quad\quad\text{for $G_1$-to-$G_2$ prediction},
		\\
		\widetilde{P}_k(\overline{z}_1,\overline{z}_2)
		&=
		\frac{W_1(1+z_1z_2)+W_2(z_1+z_2)}{W_1+W_2}
		p_k
		\nonumber
		\\
		&\qquad\qquad\qquad\quad\quad\text{for $B$-to-$R$ prediction}
		,
	\end{align}
and define $w_{1n}$ and $w_{2n}$ as
	\begin{align}
		w_{1n}
		&=
		\begin{cases}
			|Z_n(z_1-\overline{z}_2)G_1|^\gamma & \text{for $G_1$-to-$G_2$ prediction} \\
			|Z_n(z_1-z_2)B|^\gamma              & \text{for $B$-to-$R$ prediction}
		\end{cases}
		,
		\\
		w_{2n}
		&=
		\begin{cases}
			|Z_n(1-z_1\overline{z}_2)G_1|^\gamma & \text{for $G_1$-to-$G_2$ prediction} \\
			|Z_n(1-z_1z_2)B|^\gamma              & \text{for $B$-to-$R$ prediction}
		\end{cases}
		,
	\end{align}
where $\gamma\in\mathbb{R}$ is an arbitrary adjustment parameter that evaluates the edge-likeness ($\widetilde{P}_k(z_1,z_2)=P_k(z_1,z_2)$ if $\gamma=0$) and $Z_n$ is a delay:
	\begin{align}
		Z_n
		=
		\begin{cases}
			1                            & \text{if $n=0$ (center)}       \\
			z_2                          & \text{if $n=1$ (top)}          \\
			z_1                          & \text{if $n=2$ (left)}         \\
			\overline{z}_1               & \text{if $n=3$ (right)}        \\
			\overline{z}_2               & \text{if $n=4$ (bottom)}       \\
			z_1z_2                       & \text{if $n=5$ (top-left)}     \\
			\overline{z}_1z_2            & \text{if $n=6$ (top-right)}    \\
			z_1\overline{z}_2            & \text{if $n=7$ (bottom-left)}  \\
			\overline{z}_1\overline{z}_2 & \text{if $n=8$ (bottom-right)}
		\end{cases}
		.
	\end{align}
Figure~\ref{Fig_WeightsDiag} overviews the weights for the edge-aware weighted 2-D diagonal predict steps.
$\gamma$ was empirically set to $\gamma=1$ in the lossless compression experiment and $\gamma=1/2$ in the lossy compression experiment.
Since the edge-aware weighted 2-D diagonal predict steps can be applied to any $P_k(z_1,z_2)$ in \eqref{Eqs_DWT2}, we can extend all of the WSSTs and XSTTs to efficiently decorrelate the CFA-sampled raw images while taking the edge information into account.
Also, to avoid extra bits and a significant amount of complexity, the weights are not transmitted to the decoder and are re-calculated in the decoder.
Note that, as in Section~\ref{Sec_RawImageCompression}, the weights between the encoder and decoder tend to be more different at higher lossy compression levels.

\subsection{Edge-Aware Weighted 2-D Horizontal-Vertical Predict Steps}
	\begin{figure}[t]
		\centering
		\includegraphics[scale=0.3,keepaspectratio=true]{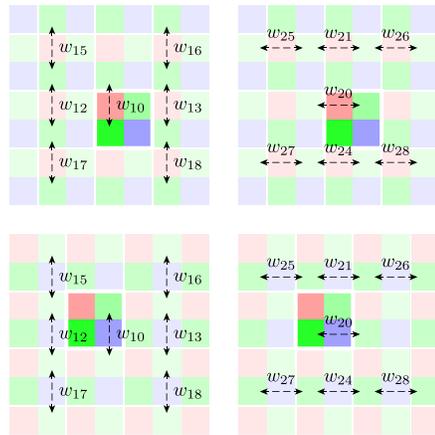}
		\caption{Weights for the 2-D horizontal-vertical predict steps (dashed arrows mean to calculate the difference between two pixels):
		(top) $(G_1,G_2)$-to-$R$ prediction
		and
		(bottom) $(G_1,G_2)$-to-$B$ prediction.}
		\label{Fig_WeightsHoriVert}
	\end{figure}

Like the {edge-aware weighted 2-D diagonal predict steps} in \eqref{Eqs_Pk_Diag}, we can also extend the polynomial $P_k(z_i)$ in \eqref{Eqs_1DP}, which is in the 2-D \textit{horizontal-vertical} predict step in \eqref{Eqs_XSTT2_PU}, as follows:
	\begin{align}
		\widetilde{P}_k(z_i)
		=
		\frac{W_i}{W_1+W_2}(1+\overline{z}_i)p_k
		.
		\label{Eqs_Pk_HoriVert}
	\end{align}
When $W_1=W_2=1$, it is clear that $\widetilde{P}_k(z_i)=P_k(z_i)$.
Naturally, we can rewrite $\widetilde{P}_k(z_i)$ in \eqref{Eqs_Pk_HoriVert} as
	\begin{align}
		\widetilde{P}_k(\overline{z}_i)
		=
		\frac{W_i}{W_1+W_2}(1+z_i)p_k
		.
	\end{align}
Moreover, we can rewrite $w_{mn}$ as
	\begin{align}
		w_{1n}
		&=
		\begin{cases}
			|Z_n(1-z_2)G_2|^\gamma            & \text{for $(G_1,G_2)$-to-$R$ prediction} \\
			|Z_n(1-\overline{z}_2)G_1|^\gamma & \text{for $(G_1,G_2)$-to-$B$ prediction}
		\end{cases}
		,
		\\
		w_{2n}
		&=
		\begin{cases}
			|Z_n(1-z_1)G_1|^\gamma            & \text{for $(G_1,G_2)$-to-$R$ prediction} \\
			|Z_n(1-\overline{z}_1)G_2|^\gamma & \text{for $(G_1,G_2)$-to-$B$ prediction}
		\end{cases}
		,
	\end{align}
where $\gamma$ and $Z_n$ are the same as in the case of the edge-aware weighted 2-D diagonal predict steps.
Figure~\ref{Fig_WeightsHoriVert} overviews the weights for the 2-D horizontal-vertical predict steps.
Since the vertical (horizontal) weight $w_{mn}$ detects the horizontal (vertical) edge direction with a $90^\circ$ range, the top and bottom (left and right) weights are omitted.
Like the edge-aware weighted 2-D diagonal predict steps, $\gamma$ was empirically set as $\gamma=1$ in the lossless compression experiment and $\gamma=1/2$ in the lossy compression experiment.
Since the edge-aware weighted 2-D horizontal-vertical predict steps can be applied to only $P_k(z_i)$ in \eqref{Eqs_XSTT1_PU}, unlike the diagonal case, and $P_k(z_i)$ in \eqref{Eqs_XSTT1_PU} is included only in the XSTTs, we will be able to extend only the XSTTs.
Also, as in the diagonal case, the weights are re-calculated in the decoder.

\section{Experimental Results}\label{Sec_Result}
\subsection{Preparation}
	\begin{figure*}[t]
		\centering
		\includegraphics[scale=0.7,keepaspectratio=true]{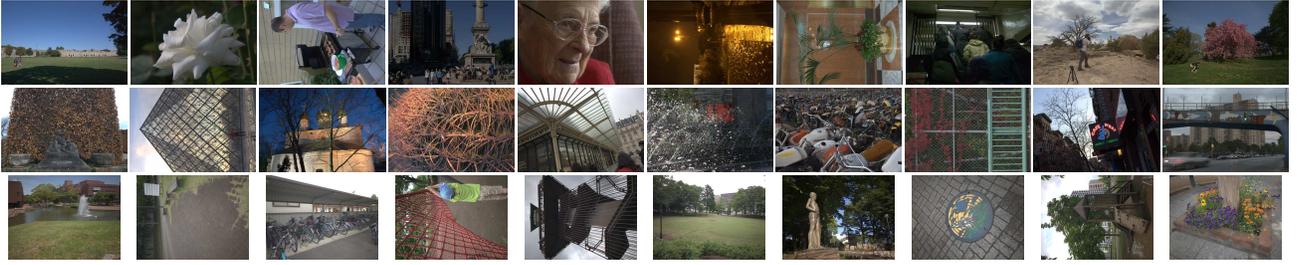}
		\caption{Test images developed from raw data by a simple camera processing pipeline: (top) general images (\textit{\#0500}, \textit{\#1000}, \textit{\#1500}, \textit{\#2000}, \textit{\#2500}, \textit{\#3000}, \textit{\#3500}, \textit{\#4000}, \textit{\#4500}, and \textit{\#5000}), (middle) images with many edges (\textit{\#0092}, \textit{\#0482}, \textit{\#0548}, \textit{\#1137}, \textit{\#1145}, \textit{\#1463}, \textit{\#2239}, \textit{\#2912}, \textit{\#3394}, and \textit{\#3419}), and (bottom) mobile phone images (\textit{m01}, \textit{m02}, \textit{m03}, \textit{m04}, \textit{m05}, \textit{m06}, \textit{m07}, \textit{m08}, \textit{m09}, and \textit{m10}).}
		\label{Fig_Images}
	\end{figure*}

We compared our XSTTs and EXSTTs, which are the XSTTs that use the edge-aware weighted 2-D predict steps, with existing spectral-spatial transforms including simple ones which do not apply white balance or gamma correction to themselves for CAMRA~\cite{Lee2018TIP}, the WSSTs~\cite{Suzuki2020TIP}, the STT~\cite{Richter2021TIP} which do not apply white-balancing constants to themselves, and the weighted WSSTs (WWSSTs)~\cite{Huang2022ICASSP}.\footnote{The YDgCoCg-WSSTs/WWSSTs performed the best among the three types of WSSTs/WWSSTs in preliminary experiments, so the comparisons presented here show only YDgCoCg-WSSTs/WWSSTs; hereafter, the YDgCoCg-WSSTs/WWSSTs will simply be referred to as ``WSSTs/WWSSTs.'' In addition, the weights of WWSST were updated to the ones in this study.}
For reference, we compared them with direct JPEG 2000 compression of full-color RGB images obtained by a simple camera processing pipeline (black level correction, white balance, demosaicing~\cite{Malvar2004ICASSP}, and gamma correction), as shown in \cite{ImplDigCamProcPip}.
The 5/3 and 9/7 wavelet transforms\footnote{We did not apply rounding operations to the 9/7 wavelet transforms for better lossy compression.} were used in the lossless and lossy compressions based on the spectral-spatial transforms; the XSTT-II and EXSTT-II have 5/3 wavelet transforms for the first and last steps even for lossy compression because of their special structures.
Note that the XSTT-I that uses 5/3 wavelet transforms is identical to the STT and the one that uses 9/7 wavelet transforms can be regarded as an extended version of the STT, as indicated above.
The transformed images were adjusted to positive values and compressed with JPEG 2000~\cite{Skodras2001IEEE}\footnote{JPEG 2000 that uses the wavelet transforms has high affinity to any spectral-spatial transform. Of course, other codecs, such as JPEG XR and JPEG XS, can be applied to any of the spectral-spatial transforms.} by using `imwrite.m' with the default settings (except for compression ratio) in MATLAB.
To investigate the effect on high-quality camera images, we used 20 DNG data, whose bit depths are $12$ or $14$ bits, in the MIT-Adobe FiveK dataset~\cite{Bychkovsky2011CVPR} after loading them via `rawread.m' in MATLAB and forming them into an ``RGGB'' array.
The high-quality camera images included ten general images used in \cite{Suzuki2020TIP} and ten intuitively selected images that seemed to have many edges.
To investigate the effect on mobile phone images, which are noisier than the images in the MIT-Adobe FiveK dataset, we used ten DNG data, whose bit depths are 12 bits, acquired with an iPhone SE (3rd generation).
Figure~\ref{Fig_Images} shows the 30 test images.
In addition, we used the lossless bitrates (LBRs) [bpp] in lossless compression and the Bj\o ntegaard delta peak signal-to-noise ratios (BD-PSNRs) [dB] between about $2$--$5$ bpp in lossy compression for a fair comparison.
We selected the macropixel spectral-spatial transform (MSST)~\cite{Malvar2012DCC} as the basis for the comparison using the BD-PSNRs and computed the BD-PSNRs after the CFA-sampled images were developed into full-color RGB images by using a simple camera processing pipeline similar to that used for direct JPEG 2000 compression.

\subsection{Decorrelation of CFA-Sampled Raw Images}
	\begin{figure}[t]
		\centering
		\includegraphics[scale=0.5,keepaspectratio=true]{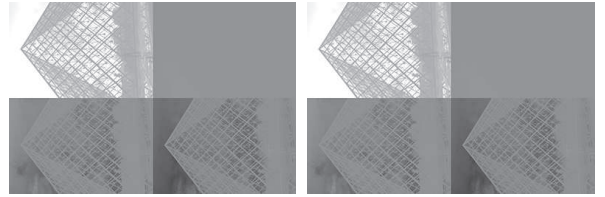}
		\caption{Decorrelated subband images of \textit{\#0482} (clockwise from the top-left: $Y$, $D_\text{g}$, $C_2$, and $C_1$): (left) XSTT-I and (right) EXSTT-I that use 9/7 wavelet transforms.}
		\label{Fig_Decorrelation}
	\end{figure}
	\begin{figure}[t]
		\centering
		\includegraphics[scale=0.4,keepaspectratio=true]{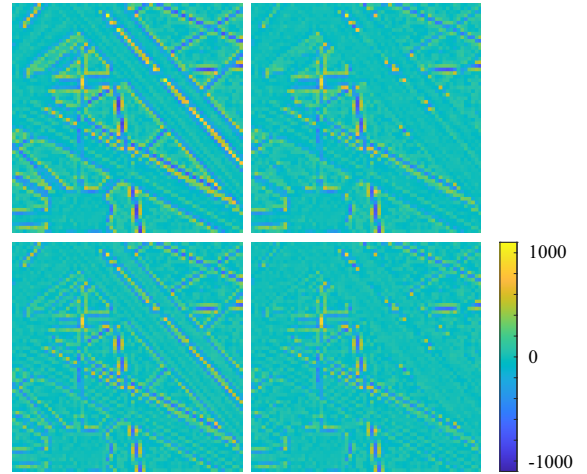}
		\caption{Part of $D_\text{g}$ components of \textit{\#0482} represented in pseudo color: (top) XSTT-I and EXSTT-I that use 5/3 wavelet transforms, (bottom) XSTT-I and EXSTT-I that use 9/7 wavelet transforms.}
		\label{Fig_Dg}
	\end{figure}
	\begin{table*}[t]
		\centering
		\caption{Improvement in MSEs [\%] of $D_\text{g}$ components after incorporating edge-aware weighted 2-D predict steps in the XSTTs\\(lower MSEs are better).}
		\begin{tabular}{wc{7.5mm}|wc{7.5mm}wc{7.5mm}wc{7.5mm}wc{7.5mm}||wc{7.5mm}|wc{7.5mm}wc{7.5mm}wc{7.5mm}wc{7.5mm}||wc{7.5mm}|wc{7.5mm}wc{7.5mm}wc{7.5mm}wc{7.5mm}}
		\thline
		& \multicolumn{2}{c}{type-I} & \multicolumn{2}{c||}{type-II} && \multicolumn{2}{c}{type-I} & \multicolumn{2}{c||}{type-II} && \multicolumn{2}{c}{type-I} & \multicolumn{2}{c}{type-II} \\
		& 5/3 & 9/7 & 5/3 & 9/7 && 5/3 & 9/7 & 5/3 & 9/7 && 5/3 & 9/7 & 5/3 & 9/7 \\
		\hline
		\textit{\#0500} & $\tr{+1.42}$  & $\tr{+1.51}$  & $\tr{+1.19}$  & $\tr{+1.37}$ & \textit{\#0092} & $\tb{-16.17}$ & $\tb{-12.38}$ & $\tb{-14.72}$ & $\tb{-11.48}$ & \textit{m01} & $\tb{-9.21}$  & $\tb{-5.71}$  & $\tb{-8.43}$  & $\tb{-5.01}$  \\
		\textit{\#1000} & $\tr{+1.24}$  & $\tr{+1.40}$  & $\tr{+0.36}$  & $\tr{+1.27}$ & \textit{\#0482} & $\tb{-41.66}$ & $\tb{-34.05}$ & $\tb{-39.25}$ & $\tb{-31.76}$ & \textit{m02} & $\tb{-6.66}$  & $\tb{-3.64}$  & $\tb{-6.18}$  & $\tb{-3.17}$  \\
		\textit{\#1500} & $\tb{-17.00}$ & $\tb{-10.91}$ & $\tb{-13.59}$ & $\tb{-9.29}$ & \textit{\#0548} & $\tb{-17.04}$ & $\tb{-10.95}$ & $\tb{-14.85}$ & $\tb{-9.83}$  & \textit{m03} & $\tb{-20.58}$ & $\tb{-16.22}$ & $\tb{-15.78}$ & $\tb{-12.72}$ \\
		\textit{\#2000} & $\tb{-13.17}$ & $\tb{-8.86}$  & $\tb{-9.03}$  & $\tb{-5.49}$ & \textit{\#1137} & $\tb{-31.92}$ & $\tb{-23.62}$ & $\tb{-30.20}$ & $\tb{-22.74}$ & \textit{m04} & $\tb{-17.42}$ & $\tb{-11.34}$ & $\tb{-16.59}$ & $\tb{-10.83}$ \\
		\textit{\#2500} & $\tr{+0.66}$  & $\tr{+0.68}$  & $\tr{+0.41}$  & $\tr{+0.73}$ & \textit{\#1145} & $\tb{-26.42}$ & $\tb{-19.31}$ & $\tb{-26.14}$ & $\tb{-18.68}$ & \textit{m05} & $\tb{-29.04}$ & $\tb{-23.56}$ & $\tb{-24.50}$ & $\tb{-19.60}$ \\
		\textit{\#3000} & $\tb{-4.89}$  & $\tb{-2.78}$  & $\tb{-4.29}$  & $\tb{-1.99}$ & \textit{\#1463} & $\tb{-17.91}$ & $\tb{-14.20}$ & $\tb{-17.35}$ & $\tb{-14.52}$ & \textit{m06} & $\tb{-11.40}$ & $\tb{-7.72}$  & $\tb{-10.24}$ & $\tb{-6.94}$  \\
		\textit{\#3500} & $\tb{-16.09}$ & $\tb{-7.09}$  & $\tb{-14.83}$ & $\tb{-5.91}$ & \textit{\#2239} & $\tb{-26.81}$ & $\tb{-19.08}$ & $\tb{-25.11}$ & $\tb{-18.24}$ & \textit{m07} & $\tb{-10.03}$ & $\tb{-6.22}$  & $\tb{-9.34}$  & $\tb{-5.72}$  \\
		\textit{\#4000} & $\tb{-8.22}$  & $\tb{-8.53}$  & $\tb{-1.82}$  & $\tb{-1.17}$ & \textit{\#2912} & $\tb{-25.99}$ & $\tb{-12.83}$ & $\tb{-24.51}$ & $\tb{-10.31}$ & \textit{m08} & $\tb{-7.87}$  & $\tb{-3.79}$  & $\tb{-8.13}$  & $\tb{-4.16}$  \\
		\textit{\#4500} & $\tb{-13.33}$ & $\tb{-9.25}$  & $\tb{-11.80}$ & $\tb{-7.82}$ & \textit{\#3394} & $\tb{-25.95}$ & $\tb{-20.36}$ & $\tb{-23.04}$ & $\tb{-18.22}$ & \textit{m09} & $\tb{-22.69}$ & $\tb{-18.72}$ & $\tb{-19.96}$ & $\tb{-16.67}$ \\
		\textit{\#5000} & $\tb{-6.67}$  & $\tb{-3.32}$  & $\tb{-6.59}$  & $\tb{-3.25}$ & \textit{\#3419} & $\tb{-41.44}$ & $\tb{-38.39}$ & $\tb{-38.31}$ & $\tb{-36.41}$ & \textit{m10} & $\tb{-9.25}$  & $\tb{-4.89}$  & $\tb{-9.22}$  & $\tb{-5.07}$  \\
		\rowcolor[gray]{0.8}%
		Avg.            & $\tb{-9.84}$  & $\tb{-7.66}$  & $\tb{-5.92}$  & $\tb{-3.38}$ & Avg.            & $\tb{-30.08}$ & $\tb{-24.82}$ & $\tb{-27.81}$ & $\tb{-23.23}$ & Avg.         & $\tb{-14.47}$ & $\tb{-10.26}$ & $\tb{-12.81}$ & $\tb{-8.97}$  \\
		\thline
		\end{tabular}
		\label{Tab_MSE_Dg}
	\end{table*}
	\begin{figure*}[t]
		\centering
		\includegraphics[scale=0.4,keepaspectratio=true]{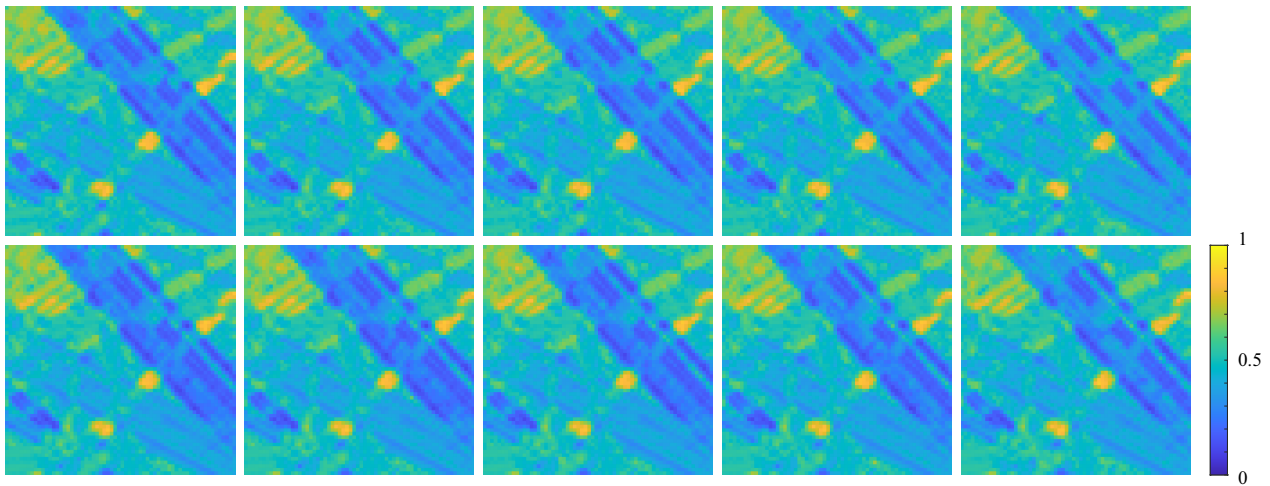}
		\caption{Weights for $G1$-to-$G2$ predictions, which generate $D_\text{g}$ components, in the EXSTT-Is for part of \textit{\#0482} represented in pseudo color: (top) $W_1/(W_1+W_2)$ in 5/3 wavelet transforms, (bottom) the first $W_1/(W_1+W_2)$ in 9/7 wavelet transforms, (first column) ideal weights on the encoder, and (second-to-last columns) lossy weights on the decoder when 5, 4, 3, and $2$ bpp.}
		\label{Fig_WeightsImage}
	\end{figure*}
Figure~\ref{Fig_Decorrelation} shows only the decorrelated (transformed) subband images with the XSTT-I and EXSTT-I that use 9/7 wavelet transforms because the differences depending on the transforms were not clear.
The grayscale-transformed images were directly compressed with JPEG 2000.

To clearly show the effect of the edge-aware weighted 2-D predict steps, Fig.~\ref{Fig_Dg} and Table~\ref{Tab_MSE_Dg} show parts of the $D_\text{g}$ components, the high-frequency information between the $G_1$ and $G_2$ components, of the images transformed by the XSTTs/EXSTTs and the reduction (improvement) in mean squared error (MSE)\footnote{Strictly speaking, these values are not MSEs. However, we indicate them as such because $D_\text{g}$ is like the error between the $G_1$ and $G_2$ components.} [\%] of the $Dg$ components after applying the edge-aware weighted 2-D predict steps to the XSTTs.
Since the $D_\text{g}$ components in the XSTT-II/EXSTT-IIs look almost identical to the ones in the XSTT-I/EXSTT-Is, we omitted the XSTT-II/EXSTT-IIs from Fig.~\ref{Fig_Dg}.
The EXSTTs reduced the average MSEs of the $D_\text{g}$ components by about $3.38$--$9.84$ \% for general images, $23.23$--$30.08$ \% for images with many edges, and $8.97$--$14.47$ \% for mobile phone images.
Figure~\ref{Fig_WeightsImage} shows the corresponding weights for the part in Fig.~\ref{Fig_Dg}.
We omitted the second weights in the 9/7 wavelet transforms, which looked almost identical to the first weights.
Figures~\ref{Fig_Dg} and \ref{Fig_WeightsImage} show that the weights assigned along the edge directions helped to reduce the $D_\text{g}$ energies.
There were also slight differences in weight between the 5/3 and 9/7 wavelet transforms.
On the other hand, unfortunately, the weights between the encoder and decoder tended to be more different for higher lossy compression levels.
Since the differences may affect coding performance, we will investigate this finding in Section~\ref{Sec_RawImageCompression}.

\subsection{CFA-Sampled Raw Image Compression}\label{Sec_RawImageCompression}
\subsubsection{Effect of Edge-Aware Weighted 2-D Predict Steps}
	\begin{table*}[t]
		\centering
		\caption{Improvement in LBRs [bpp] for lossless compression after incorporating edge-aware weighted 2-D predict steps in the XSTTs (lower LBRs are better).}
		\begin{tabular}{c|cc||c|cc||c|cc}
		\thline
		& type-I & type-II && type-I & type-II && type-I & type-II \\
		\hline
		\textit{\#0500} & $\tr{+0.01}$ & $\tr{+0.01}$ & \textit{\#0092} & $\tb{-0.02}$ & $\tb{-0.02}$ & \textit{m01} & $\pm 0.00$   & $\pm 0.00$   \\
		\textit{\#1000} & $\tr{+0.01}$ & $\tr{+0.01}$ & \textit{\#0482} & $\tb{-0.08}$ & $\tb{-0.08}$ & \textit{m02} & $\pm 0.00$   & $\pm 0.00$   \\
		\textit{\#1500} & $\pm 0.00$   & $\pm 0.00$   & \textit{\#0548} & $\tb{-0.02}$ & $\tb{-0.02}$ & \textit{m03} & $\tb{-0.01}$ & $\tb{-0.01}$ \\
		\textit{\#2000} & $\tr{+0.01}$ & $\tr{+0.01}$ & \textit{\#1137} & $\tb{-0.02}$ & $\tb{-0.02}$ & \textit{m04} & $\tb{-0.01}$ & $\tb{-0.01}$ \\
		\textit{\#2500} & $\tr{+0.01}$ & $\tr{+0.01}$ & \textit{\#1145} & $\tb{-0.02}$ & $\tb{-0.02}$ & \textit{m05} & $\tb{-0.01}$ & $\tb{-0.01}$ \\
		\textit{\#3000} & $\pm 0.00$   & $\pm 0.00$   & \textit{\#1463} & $\tb{-0.02}$ & $\tb{-0.02}$ & \textit{m06} & $\tb{-0.01}$ & $\tb{-0.01}$ \\
		\textit{\#3500} & $\pm 0.00$   & $\pm 0.00$   & \textit{\#2239} & $\tb{-0.02}$ & $\tb{-0.02}$ & \textit{m07} & $\tb{-0.01}$ & $\tb{-0.01}$ \\
		\textit{\#4000} & $\tr{+0.01}$ & $\tr{+0.01}$ & \textit{\#2912} & $\tb{-0.04}$ & $\tb{-0.04}$ & \textit{m08} & $\tb{-0.01}$ & $\tb{-0.01}$ \\
		\textit{\#4500} & $\pm 0.00$   & $\pm 0.00$   & \textit{\#3394} & $\tb{-0.02}$ & $\tb{-0.02}$ & \textit{m09} & $\tb{-0.01}$ & $\tb{-0.01}$ \\
		\textit{\#5000} & $\pm 0.00$   & $\pm 0.00$   & \textit{\#3419} & $\tb{-0.02}$ & $\tb{-0.02}$ & \textit{m10} & $\tb{-0.01}$ & $\tb{-0.01}$ \\
		\rowcolor[gray]{0.8}%
		Avg.            & $\pm 0.00$   & $\pm 0.00$   & Avg.            & $\tb{-0.03}$ & $\tb{-0.03}$ & Avg.         & $\tb{-0.01}$ & $\tb{-0.01}$ \\
		\thline
		\end{tabular}
		\label{Tab_ImprovedResultLossless}
	\end{table*}
	\begin{table*}[t]
		\centering
		\caption{Improvement in BD-PSNRs [dB] for lossy compression after incorporating edge-aware weighted 2-D predict steps in the XSTTs (higher BD-PSNRs are better, and the values in () indicate the ideal weights without side information considered).}
		\begin{tabular}{wc{7.5mm}|wc{7.5mm}wc{7.5mm}wc{7.5mm}wc{7.5mm}||wc{7.5mm}|wc{7.5mm}wc{7.5mm}wc{7.5mm}wc{7.5mm}||wc{7.5mm}|wc{7.5mm}wc{7.5mm}wc{7.5mm}wc{7.5mm}}
		\thline
		& \multicolumn{2}{c}{type-I} & \multicolumn{2}{c||}{type-II} && \multicolumn{2}{c}{type-I} & \multicolumn{2}{c||}{type-II} && \multicolumn{2}{c}{type-I} & \multicolumn{2}{c}{type-II} \\
		\hline
		\textit{\#0500} & $\tr{-0.27}$ & ($\tr{-0.06}$) & $\tr{-0.15}$ & ($\tr{-0.07}$) & \textit{\#0092} & $\tr{-0.62}$ & ($\tb{+0.10}$) & $\tr{-1.54}$ & ($\tb{+0.05}$) & \textit{m01} & $\tr{-0.29}$ & ($\tr{-0.04}$) & $\tr{-0.09}$ & ($\tr{-0.04}$) \\
		\textit{\#1000} & $\tr{-0.36}$ & ($\tr{-0.11}$) & $\tr{-0.15}$ & ($\tr{-0.05}$) & \textit{\#0482} & $\tr{-0.24}$ & ($\tb{+0.19}$) & $\tr{-0.08}$ & ($\tb{+0.25}$) & \textit{m02} & $\tr{-0.11}$ & ($\pm 0.00$)   & $\tr{-0.07}$ & ($\tr{-0.02}$) \\
		\textit{\#1500} & $\tr{-0.42}$ & ($\tr{-0.03}$) & $\tr{-0.78}$ & ($\tr{-0.03}$) & \textit{\#0548} & $\tr{-0.06}$ & ($\pm 0.00$)   & $\pm 0.00$   & ($\tb{+0.02}$) & \textit{m03} & $\tr{-1.59}$ & ($\tr{-0.15}$) & $\tr{-1.39}$ & ($\tr{-0.01}$) \\
		\textit{\#2000} & $\tr{-0.30}$ & ($\tr{-0.08}$) & $\tr{-0.09}$ & ($\tr{-0.01}$) & \textit{\#1137} & $\tr{-0.08}$ & ($\tb{+0.01}$) & $\tb{+0.01}$ & ($\tb{+0.05}$) & \textit{m04} & $\tr{-0.15}$ & ($\tr{-0.05}$) & $\tr{-2.82}$ & ($\tb{+0.01}$) \\
		\textit{\#2500} & $\tr{-0.33}$ & ($\tr{-0.11}$) & $\tr{-0.15}$ & ($\tr{-0.06}$) & \textit{\#1145} & $\tr{-0.06}$ & ($\tb{+0.03}$) & $\tb{+0.01}$ & ($\tb{+0.07}$) & \textit{m05} & $\tr{-0.10}$ & ($\pm 0.00$)   & $\tr{-0.41}$ & ($\tb{+0.01}$) \\
		\textit{\#3000} & $\tr{-0.82}$ & ($\tr{-0.04}$) & $\tr{-0.64}$ & ($\tb{+0.01}$) & \textit{\#1463} & $\tr{-0.21}$ & ($\tb{+0.07}$) & $\tr{-0.12}$ & ($\tb{+0.09}$) & \textit{m06} & $\tr{-0.13}$ & ($\tb{+0.05}$) & $\tr{-1.38}$ & ($\tb{+0.01}$) \\
		\textit{\#3500} & $\tr{-0.52}$ & ($\tr{-0.11}$) & $\tr{-0.10}$ & ($\tr{-0.05}$) & \textit{\#2239} & $\tr{-0.03}$ & ($\tb{+0.05}$) & $\tr{-0.04}$ & ($\tb{+0.03}$) & \textit{m07} & $\tr{-0.07}$ & ($\pm 0.00$    & $\tr{-0.07}$ & ($\tr{-0.01}$) \\
		\textit{\#4000} & $\tr{-0.78}$ & ($\tr{-0.10}$) & $\tr{-0.23}$ & ($\tr{-0.04}$) & \textit{\#2912} & $\tr{-0.03}$ & ($\tb{+0.08}$) & $\tb{+0.04}$ & ($\tb{+0.09}$) & \textit{m08} & $\tr{-0.09}$ & ($\tr{-0.01}$) & $\tr{-0.06}$ & ($\tr{-0.03}$) \\
		\textit{\#4500} & $\tr{-0.15}$ & ($\tr{-0.02}$) & $\tr{-0.06}$ & ($\pm 0.00$)   & \textit{\#3394} & $\tr{-2.13}$ & ($\tb{+0.03}$) & $\tr{-0.51}$ & ($\tb{+0.04}$) & \textit{m09} & $\tr{-0.25}$ & ($\pm 0.00$    & $\tr{-1.16}$ & ($\tr{-0.02}$) \\
		\textit{\#5000} & $\tr{-0.16}$ & ($\tr{-0.04}$) & $\tr{-0.07}$ & ($\tr{-0.02}$) & \textit{\#3419} & $\tr{-0.67}$ & ($\tr{-0.01}$) & $\tr{-0.09}$ & ($\tb{+0.08}$) & \textit{m10} & $\tr{-0.04}$ & ($\tb{+0.04}$) & $\tb{+0.01}$ & ($\tb{+0.05}$) \\
		\rowcolor[gray]{0.8}%
		Avg.            & $\tr{-0.41}$ & ($\tr{-0.07}$) & $\tr{-0.24}$ & ($\tr{-0.03}$) & Avg.            & $\tr{-0.41}$ & ($\tb{+0.06}$) & $\tr{-0.23}$ & ($\tb{+0.08}$) & Avg.         & $\tr{-0.28}$ & ($\tr{-0.01}$) & $\tr{-0.74}$ & ($\tr{-0.01}$) \\
		\thline
		\end{tabular}
		\label{Tab_ImprovedResultLossy}
	\end{table*}

Tables~\ref{Tab_ImprovedResultLossless} and \ref{Tab_ImprovedResultLossy} show the improvements in the lossless and lossy compressions for the CFA-sampled raw images after incorporating the edge-aware weighted 2-D predict steps in the XSTTs: the differences in LBR and BD-PSNR were obtained by subtracting the LBR and BD-PSNR of the original transforms from those of the edge-aware transforms in lossless and lossy compression, respectively.
For lossless compression, although all EXSTTs hardly showed any differences on the general and mobile phone images, they gave the best results in the case of images with many edges.
In contrast, for lossy compression, the EXSTTs gave worse results than the XSTTs because we re-calculated \textit{lossy} weights on the decoder (Fig.~\ref{Fig_WeightsImage}).
If we would like to use the \textit{ideal} weights on the decoder, the encoder-calculated weights must be transmitted to the decoder together with a lot of side information and it is more practical to use the XSTTs instead of the EXSTTs for lossy compression.

\subsubsection{Comparisons with Other Methods}
	\begin{table*}[t]
		\centering
		\caption{LBRs [bpp] in lossless compression.}
		\begin{tabular}{c|c|ccccc|cccc}
		\thline
		& Direct & MSST~\cite{Malvar2012DCC} & CAMRA~\cite{Lee2018TIP} & WSST~\cite{Suzuki2020TIP} & STT~\cite{Richter2021TIP} & WWSST~\cite{Huang2022ICASSP} & XSTT-I & XSTT-II & EXSTT-I & EXSTT-II \\
		\hline
		\textit{\#0500} & ($37.56$) & $6.43$      & $6.41$ & $\mb{6.39}$ & $6.40$      & $6.40$      & $6.40$      & $6.42$ & $6.41$      & $6.43$ \\
		\textit{\#1000} & ($34.84$) & $6.38$      & $6.40$ & $\mb{6.37}$ & $6.40$      & $6.38$      & $6.40$      & $6.41$ & $6.41$      & $6.42$ \\
		\textit{\#1500} & ($33.39$) & $8.44$      & $8.38$ & $\mb{8.32}$ & $\mb{8.32}$ & $\mb{8.32}$ & $\mb{8.32}$ & $8.34$ & $\mb{8.32}$ & $8.34$ \\
		\textit{\#2000} & ($34.14$) & $5.27$      & $5.25$ & $5.23$      & $\mb{5.19}$ & $5.22$      & $\mb{5.19}$ & $5.20$ & $\mb{5.19}$ & $5.21$ \\
		\textit{\#2500} & ($36.24$) & $6.27$      & $6.31$ & $\mb{6.25}$ & $6.28$      & $\mb{6.25}$ & $6.28$      & $6.30$ & $6.29$      & $6.30$ \\
		\textit{\#3000} & ($37.94$) & $5.98$      & $5.97$ & $5.92$      & $\mb{5.91}$ & $5.93$      & $\mb{5.91}$ & $5.92$ & $\mb{5.91}$ & $5.92$ \\
		\textit{\#3500} & ($29.80$) & $7.46$      & $7.42$ & $7.36$      & $\mb{7.35}$ & $7.36$      & $\mb{7.35}$ & $7.37$ & $\mb{7.35}$ & $7.37$ \\
		\textit{\#4000} & ($39.72$) & $\mb{8.62}$ & $8.66$ & $\mb{8.62}$ & $8.65$      & $\mb{8.62}$ & $8.65$      & $8.66$ & $8.66$      & $8.67$ \\
		\textit{\#4500} & ($34.41$) & $7.12$      & $7.04$ & $7.02$      & $6.94$      & $7.00$      & $6.94$      & $6.95$ & $\mb{6.93}$ & $6.94$ \\
		\textit{\#5000} & ($35.30$) & $6.14$      & $6.11$ & $\mb{5.98}$ & $6.01$      & $\mb{5.98}$ & $6.01$      & $6.03$ & $6.01$      & $6.03$ \\
		\rowcolor[gray]{0.8}%
		Avg.            & ($35.33$) & $6.81$      & $6.80$ & $\mb{6.75}$ & $\mb{6.75}$ & $\mb{6.75}$ & $\mb{6.75}$ & $6.76$ & $\mb{6.75}$ & $6.76$ \\
		\hline
		\textit{\#0092} & ($34.03$) & $7.21$      & $7.15$ & $7.05$ & $6.88$ & $7.01$ & $6.88$ & $6.85$ & $6.86$      & $\mb{6.83}$ \\
		\textit{\#0482} & ($34.17$) & $9.19$      & $9.07$ & $9.05$ & $9.01$ & $8.95$ & $9.01$ & $9.01$ & $\mb{8.93}$ & $\mb{8.93}$ \\
		\textit{\#0548} & ($35.69$) & $6.98$      & $6.81$ & $6.84$ & $6.77$ & $6.83$ & $6.77$ & $6.78$ & $\mb{6.75}$ & $6.76$      \\
		\textit{\#1137} & ($34.06$) & $8.31$      & $8.22$ & $8.22$ & $8.17$ & $8.19$ & $8.17$ & $8.18$ & $\mb{8.15}$ & $8.16$      \\
		\textit{\#1145} & ($34.02$) & $8.96$      & $8.93$ & $8.85$ & $8.83$ & $8.82$ & $8.83$ & $8.84$ & $\mb{8.81}$ & $\mb{8.81}$ \\
		\textit{\#1463} & ($37.46$) & $\mb{7.46}$ & $7.58$ & $7.53$ & $7.48$ & $7.51$ & $7.48$ & $7.48$ & $\mb{7.46}$ & $7.47$      \\
		\textit{\#2239} & ($33.99$) & $6.30$      & $6.23$ & $6.21$ & $6.15$ & $6.19$ & $6.15$ & $6.17$ & $\mb{6.13}$ & $6.15$      \\
		\textit{\#2912} & ($34.04$) & $6.16$      & $6.06$ & $5.99$ & $5.96$ & $5.95$ & $5.96$ & $5.98$ & $\mb{5.91}$ & $5.93$      \\
		\textit{\#3394} & ($36.00$) & $8.50$      & $8.45$ & $8.40$ & $8.37$ & $8.39$ & $8.37$ & $8.37$ & $\mb{8.36}$ & $\mb{8.36}$ \\
		\textit{\#3419} & ($32.11$) & $8.29$      & $8.19$ & $8.15$ & $8.08$ & $8.13$ & $8.08$ & $8.09$ & $\mb{8.06}$ & $8.07$      \\
		\rowcolor[gray]{0.8}%
		Avg.            & ($34.56$) & $7.73$     & $7.67$ & $7.63$ & $7.57$ & $7.60$ & $7.57$ & $7.58$ & $\mb{7.54}$ & $7.55$      \\
        \hline
		\textit{m01} & ($36.72$) & $7.56$ & $7.49$ & $7.44$ & $7.41$      & $7.42$      & $7.41$      & $7.43$ & $\mb{7.40}$ & $7.43$      \\
		\textit{m02} & ($35.87$) & $7.44$ & $7.36$ & $7.32$ & $\mb{7.27}$ & $7.30$      & $\mb{7.27}$ & $7.30$ & $\mb{7.27}$ & $7.29$      \\
		\textit{m03} & ($30.83$) & $6.61$ & $6.69$ & $6.50$ & $6.53$      & $\mb{6.48}$ & $6.53$      & $6.49$ & $6.53$      & $\mb{6.48}$ \\
		\textit{m04} & ($35.27$) & $7.10$ & $6.99$ & $6.96$ & $6.94$      & $6.94$      & $6.94$      & $6.96$ & $\mb{6.93}$ & $6.95$      \\
		\textit{m05} & ($34.61$) & $6.23$ & $6.21$ & $6.16$ & $6.19$      & $\mb{6.13}$ & $6.19$      & $6.21$ & $6.17$      & $6.19$      \\
		\textit{m06} & ($31.71$) & $7.13$ & $7.28$ & $6.92$ & $7.00$      & $6.90$      & $7.00$      & $6.90$ & $6.99$      & $\mb{6.89}$ \\
		\textit{m07} & ($36.96$) & $6.45$ & $6.38$ & $6.30$ & $6.30$      & $\mb{6.28}$ & $6.30$      & $6.33$ & $6.29$      & $6.32$      \\
		\textit{m08} & ($36.98$) & $8.35$ & $8.21$ & $8.14$ & $\mb{8.09}$ & $8.12$      & $\mb{8.09}$ & $8.12$ & $\mb{8.09}$ & $8.11$      \\
		\textit{m09} & ($31.36$) & $6.58$ & $6.74$ & $6.44$ & $6.57$      & $\mb{6.42}$ & $6.57$      & $6.45$ & $6.56$      & $6.44$      \\
		\textit{m10} & ($36.23$) & $7.75$ & $7.64$ & $7.57$ & $7.55$      & $7.55$      & $7.55$      & $7.57$ & $\mb{7.54}$ & $7.56$      \\
		\rowcolor[gray]{0.8}%
		Avg.         & ($34.65$) & $7.12$ & $7.10$ & $6.98$ & $6.99$      & $\mb{6.96}$ & $6.99$      & $6.98$ & $6.98$      & $6.97$      \\
		\thline
		\end{tabular}
		\label{Tab_ResultLossless}
	\end{table*}
	\begin{table*}[t]
		\centering
		\caption{BD-PSNRs [dB] in lossy compression.}
		\begin{tabular}{c|c|ccccc|cccc}
		\thline
		& Direct & MSST~\cite{Malvar2012DCC} & CAMRA~\cite{Lee2018TIP} & WSST~\cite{Suzuki2020TIP} & STT~\cite{Richter2021TIP} & WWSST~\cite{Huang2022ICASSP} & XSTT-I & XSTT-II & EXSTT-I & EXSTT-II \\
		\hline
		\textit{\#0500} & ($0.75$)  & -- & $0.37$ & $1.38$      & $1.35$ & $1.19$      & $\mb{1.48}$ & $1.24$ & $1.22$ & $1.09$ \\
		\textit{\#1000} & ($0.03$)  & -- & $0.31$ & $1.31$      & $1.07$ & $1.11$      & $\mb{1.35}$ & $1.02$ & $0.99$ & $0.88$ \\
		\textit{\#1500} & ($2.33$)  & -- & $0.91$ & $\mb{1.49}$ & $1.43$ & $0.65$      & $\mb{1.49}$ & $1.40$ & $1.08$ & $0.62$ \\
		\textit{\#2000} & ($0.52$)  & -- & $0.25$ & $1.17$      & $1.18$ & $1.09$      & $\mb{1.39}$ & $1.27$ & $1.09$ & $1.18$ \\
		\textit{\#2500} & ($-3.13$) & -- & $0.21$ & $\mb{1.23}$ & $0.67$ & $1.08$      & $0.91$      & $0.58$ & $0.57$ & $0.43$ \\
		\textit{\#3000} & ($-2.46$) & -- & $0.53$ & $1.34$      & $1.24$ & $0.00$      & $\mb{1.43}$ & $1.25$ & $0.62$ & $0.61$ \\
		\textit{\#3500} & ($-1.02$) & -- & $0.71$ & $2.01$      & $1.76$ & $\mb{2.02}$ & $1.91$      & $1.72$ & $1.39$ & $1.62$ \\
		\textit{\#4000} & ($-2.74$) & -- & $0.24$ & $1.48$      & $1.31$ & $1.26$      & $\mb{1.50}$ & $1.19$ & $0.72$ & $0.96$ \\
		\textit{\#4500} & ($0.50$)  & -- & $1.11$ & $1.67$      & $1.92$ & $1.58$      & $\mb{1.98}$ & $1.89$ & $1.83$ & $1.84$ \\
		\textit{\#5000} & ($2.07$)  & -- & $0.39$ & $\mb{1.56}$ & $1.09$ & $1.45$      & $1.26$      & $1.19$ & $1.11$ & $1.12$ \\
		\rowcolor[gray]{0.8}%
		Avg.            & ($-0.31$) & -- & $0.50$ & $1.46$      & $1.30$ & $1.14$      & $\mb{1.47}$ & $1.27$ & $1.06$ & $1.03$ \\
		\hline
		\textit{\#0092} & ($1.97$)  & -- & $1.34$  & $1.97$ & $2.93$ & $-0.21$     & $2.88$      & $\mb{3.21}$ & $2.26$  & $1.67$ \\
		\textit{\#0482} & ($1.77$)  & -- & $1.60$  & $2.31$ & $2.17$ & $\mb{2.42}$ & $2.26$      & $2.34$      & $2.02$  & $2.26$ \\
		\textit{\#0548} & ($0.46$)  & -- & $0.93$  & $1.82$ & $1.67$ & $1.74$      & $\mb{1.92}$ & $1.82$      & $1.86$  & $1.82$ \\
		\textit{\#1137} & ($-0.74$) & -- & $0.80$  & $1.39$ & $1.35$ & $1.42$      & $\mb{1.50}$ & $1.42$      & $1.43$  & $1.43$ \\
		\textit{\#1145} & ($3.08$)  & -- & $0.53$  & $1.09$ & $1.00$ & $1.08$      & $\mb{1.12}$ & $1.06$      & $1.06$  & $1.07$ \\
		\textit{\#1463} & ($0.73$)  & -- & $-0.16$ & $0.60$ & $0.82$ & $0.26$      & $0.95$      & $\mb{1.04}$ & $0.74$  & $0.92$ \\
		\textit{\#2239} & ($1.52$)  & -- & $0.71$  & $1.34$ & $1.28$ & $1.32$      & $1.53$      & $\mb{1.55}$ & $1.50$  & $1.51$ \\
		\textit{\#2912} & ($0.85$)  & -- & $1.36$  & $2.24$ & $2.08$ & $2.21$      & $\mb{2.34}$ & $2.27$      & $2.31$  & $2.31$ \\
		\textit{\#3394} & ($1.38$)  & -- & $0.80$  & $1.87$ & $1.73$ & $1.80$      & $\mb{1.97}$ & $1.88$      & $-0.16$ & $1.36$ \\
		\textit{\#3419} & ($0.86$)  & -- & $0.83$  & $1.69$ & $1.66$ & $0.40$      & $\mb{1.85}$ & $1.71$      & $1.18$  & $1.62$ \\
		\rowcolor[gray]{0.8}%
		Avg.            & ($1.19$)  & -- & $0.87$  & $1.63$ & $1.67$ & $1.24$      & $\mb{1.83}$ & $\mb{1.83}$ & $1.42$  & $1.60$ \\
        \hline
		\textit{m01} & ($0.67$)  & -- & $1.29$ & $\mb{2.13}$ & $1.99$ & $\mb{2.13}$ & $1.98$ & $2.04$ & $1.69$ & $1.95$  \\
		\textit{m02} & ($0.45$)  & -- & $1.71$ & $\mb{2.33}$ & $2.12$ & $2.27$      & $2.21$ & $2.17$ & $2.10$ & $2.11$  \\
		\textit{m03} & ($-0.55$) & -- & $0.83$ & $\mb{2.03}$ & $1.70$ & $0.21$      & $1.92$ & $1.90$ & $0.33$ & $0.51$  \\
		\textit{m04} & ($-0.03$) & -- & $1.02$ & $\mb{1.94}$ & $1.70$ & $-0.87$     & $1.92$ & $1.79$ & $1.77$ & $-1.03$ \\
		\textit{m05} & ($0.95$)  & -- & $0.33$ & $\mb{1.24}$ & $0.77$ & $0.74$      & $1.06$ & $0.88$ & $0.96$ & $0.46$  \\
		\textit{m06} & ($1.85$)  & -- & $0.75$ & $\mb{1.97}$ & $1.52$ & $0.11$      & $1.55$ & $1.71$ & $1.42$ & $0.33$  \\
		\textit{m07} & ($2.72$)  & -- & $1.08$ & $\mb{1.97}$ & $1.61$ & $1.85$      & $1.85$ & $1.75$ & $1.77$ & $1.68$  \\
		\textit{m08} & ($1.83$)  & -- & $1.66$ & $\mb{2.55}$ & $2.25$ & $2.58$      & $2.25$ & $2.46$ & $2.15$ & $2.40$  \\
		\textit{m09} & ($2.70$)  & -- & $0.28$ & $\mb{1.89}$ & $1.45$ & $0.52$      & $1.54$ & $1.81$ & $1.29$ & $0.65$  \\
		\textit{m10} & ($1.05$)  & -- & $1.37$ & $\mb{2.61}$ & $2.29$ & $2.64$      & $2.46$ & $2.45$ & $2.42$ & $2.46$  \\
		\rowcolor[gray]{0.8}%
		Avg.         & ($1.16$)  & -- & $1.03$ & $\mb{2.07}$ & $1.74$ & $1.22$      & $1.87$ & $1.90$ & $1.59$ & $1.15$  \\
		\thline
		\end{tabular}
		\label{Tab_ResultLossy}
	\end{table*}
	\begin{figure*}[t]
		\centering
		\includegraphics[scale=0.35,keepaspectratio=true]{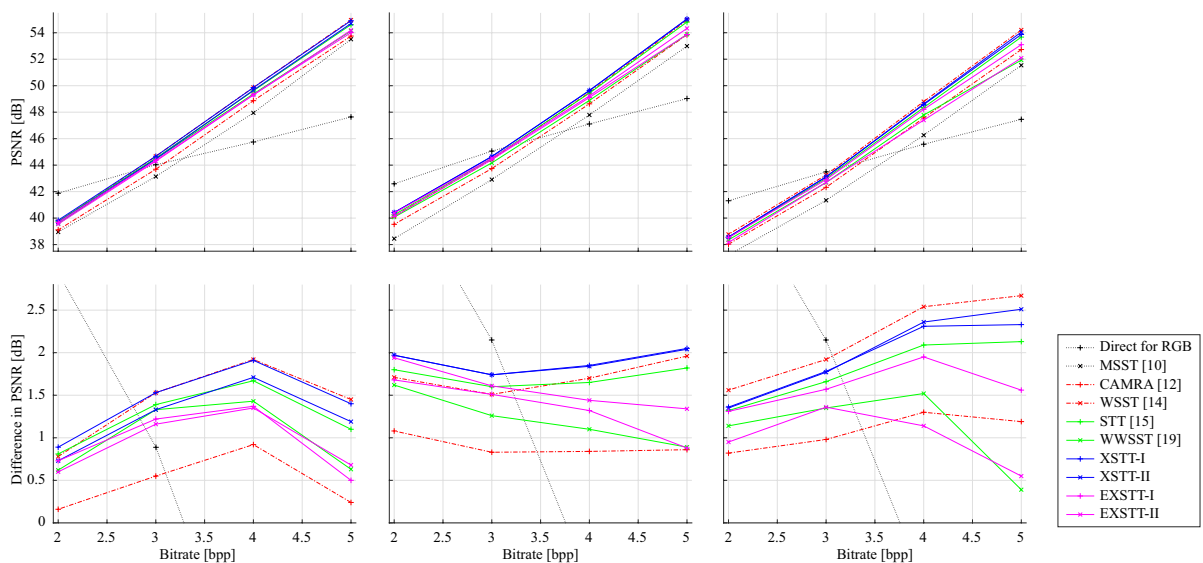}
		\caption{R-D curves of average bitrate and PSNR in lossy compression: (top) bitrate vs PSNR, (bottom) bitrate vs difference in PSNR compared with MSST (i.e., assuming that the PSNRs of the MSST~\cite{Malvar2012DCC} are $0$ dB), (left-to-right) general images, images with the many edges, and mobile phone images.}
		\label{Fig_RDcurve}
	\end{figure*}
	\begin{figure*}[t]
		\centering
		\includegraphics[scale=0.55,keepaspectratio=true]{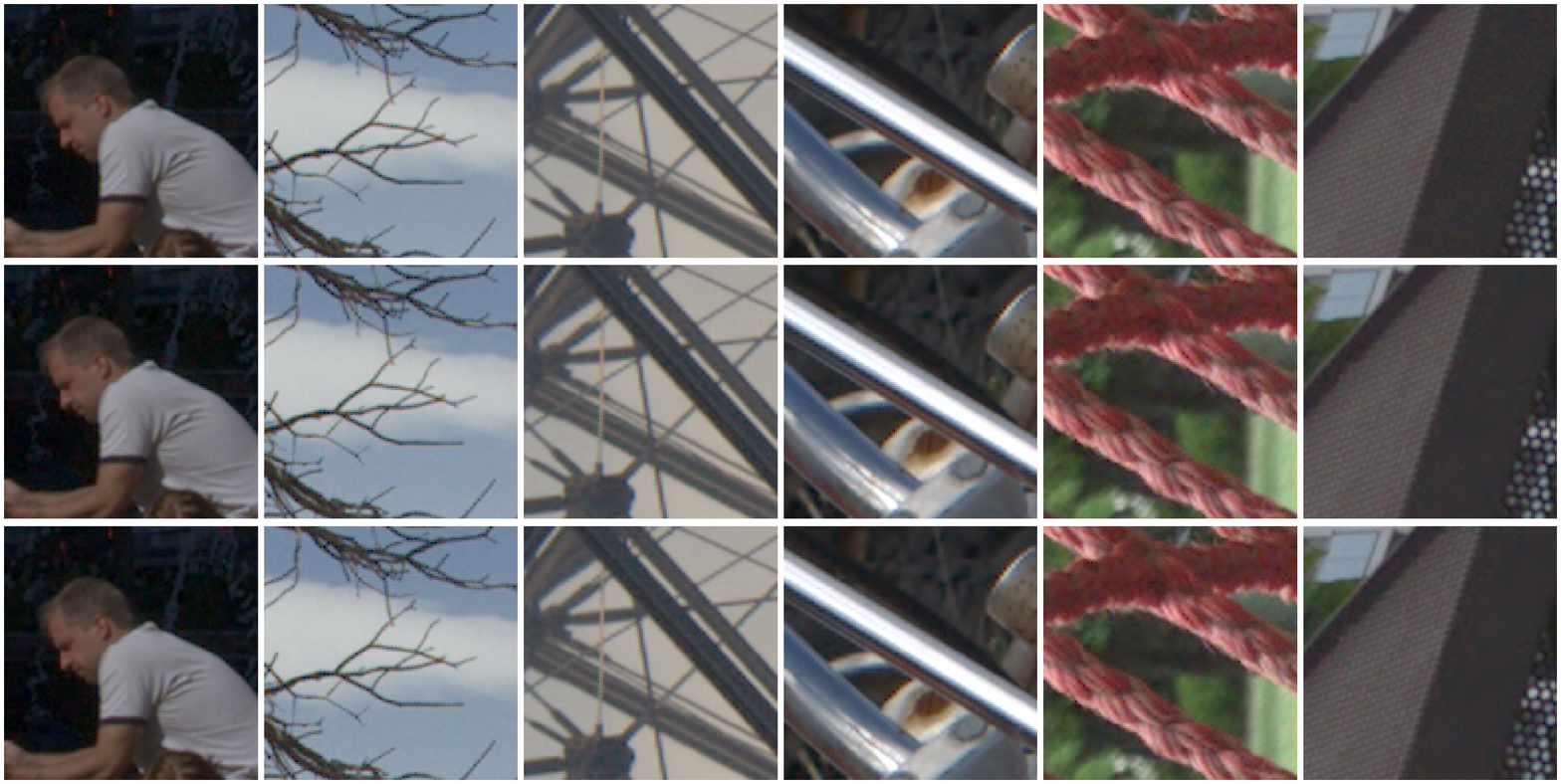}
        \caption{Part of images developed in lossy compression with $2$ bpp: (top-to-bottom) original, XSTT-I, and XSTT-II and (left-to-right) \textit{\#2000}, \textit{\#4500}, \textit{\#0482}, \textit{\#2239}, \textit{m04}, and \textit{m09}.}
		\label{Fig_DevelopImage}
	\end{figure*}

Table~\ref{Tab_ResultLossless}, Table~\ref{Tab_ResultLossy}, and Fig.~\ref{Fig_RDcurve} show the LBRs in lossless compression, BD-PSNRs in lossy compression, and rate-distortion (R-D) curves in lossy compression, respectively.
For lossless compression, the compression-first approaches using the spectral-spatial transforms clearly outperformed the direct JPEG 2000 compression of full-color RGB images, because the full-color RGB images had three times the components compared with a CFA-sampled raw image.
For lossy compression, the compression-first approaches using the spectral-spatial transforms outperformed the direct JPEG 2000 compression at higher bitrates ($4$ bits or more), but not at lower bitrates.
For high-quality camera images, the XSTTs/EXSTTs produced results equal to or better than the conventional spectral-spatial transforms: especially for images with many edges, the EXSTT-I improved them by about $0.03$--$0.19$ bpp in average LBR and the XSTTs improved them by about $0.16$--$0.96$ dB in average BD-PSNR.
In addition, for mobile phone images, i.e. noisy images, the WSSTs/WWSSTs performed the best, whereas the XSTTs/EXSTTs showed similar tends to the case of high-quality camera images.
Also, as shown in Fig.~\ref{Fig_DevelopImage}, there was no noticeable problem with the images even when they had been compressed with $2$ bpp and developed with the simple camera processing pipeline.

\section{Conclusion}\label{Sec_Concl}
This study described two types of XSTT and their edge-aware versions, EXSTTs, with no side information and little extra complexity for CFA-sampled raw image compression.
They were obtained by (i) extending the STT, which is one of the latest spectral-spatial transforms, to a new version of our previously proposed WSSTs and a simpler version, (ii) considering that each 2-D predict step of the wavelet transforms is a combination of two 1-D diagonal or horizontal-vertical transforms, and (iii) weighting the transforms along the edge directions in the images.
Compared with XSTTs, the EXSTTs decorrelated the images well.
For images with many edges, our XSTTs/EXSTTs produced results better than the conventional methods in lossless and lossy compressions without reducing the compression efficiency for general images.
For mobile phone images, our previous work performed the best, whereas the XSTTs/EXSTTs showed similar trends to the case of high-quality camera images.

\section*{Acknowledgment}
The authors would like to thank the anonymous reviewers and Dr. K. Shirai for providing many constructive suggestions that significantly improved the presentation of this paper.

\bibliographystyle{IEEEbib}
\bibliography{bib_tzszk}

\end{document}